\documentclass[conference]{IEEEtran}
\IEEEoverridecommandlockouts
\usepackage{cite}
\usepackage{amsmath,amssymb,amsfonts}
\usepackage{algorithmic}
\usepackage{graphicx}
\usepackage{textcomp}
\usepackage{xcolor}
\usepackage{booktabs}
\usepackage{stfloats}
\usepackage{url}
\def\BibTeX{{\rm B\kern-.05em{\sc i\kern-.025em b}\kern-.08em
    T\kern-.1667em\lower.7ex\hbox{E}\kern-.125emX}}
\begin{document}

\title{CXLRAMSim v1.0: System-Level Exploration of CXL Memory Expander Cards\\
}
\author{
\IEEEauthorblockN{
Karan Pathak\textsuperscript{1,2},
David Atienza\textsuperscript{1},
Marina Zapater\textsuperscript{2}
}

\IEEEauthorblockA{\textsuperscript{1}
\textit{Embedded Systems Lab, EPFL} \\
Lausanne, Switzerland \\
}

\IEEEauthorblockA{\textsuperscript{2}
\textit{REDS, School of Engineering and Management Vaud, HES-SO} \\
Switzerland \\
}
}
\maketitle

\begin{abstract}

The growing demands in the training and inference of Large Language Models (LLMs) are accelerating the adoption of scale-up systems that extend server shared memory through the use of Compute Express Link (CXL)-based load/store interconnects. Accurate full-system simulation of such architectures remains challenging, as existing tools (all very recent) rely on simplified or non-compliant architectural models, impacting accuracy and usability. 
We present CXLRAMSim, the first gem5-integrated, full-system simulator that models CXL devices at their correct position on the I/O bus, enabling the use of unmodified Linux kernels and software stack, realistic latency-bandwidth behavior and true interleaving with system DRAM. 
Our approach provides high‑fidelity CXL.mem characterization and captures key challenges such as cache pollution when accessing CXL memory.
\end{abstract}

\section{Introduction}
The rapid growth of generative AI workloads, and in particular in the training and inference of LLMs, is driving the shift toward scale‑up architectures where servers rely on CXL to extend shared memory capacity seamlessly (for instance, to distribute the KV-cache across several nodes when it does not fit a single server instance). Yet the amplification of benefits in any scaled‑up system is fundamentally bounded by the performance of its building blocks. Hence, it is essential for architects to understand these blocks to identify bottlenecks and sources of inefficiency.  
Current simulation infrastructures either oversimplify CXL, are not architecturally correct, or require modifications to the OS/software stack. 

CXLRAMSim (\textbf{CXL }D\textbf{RAM} \textbf{Sim}ulator)  fills this critical gap 
by providing a gem5‑integrated~\cite{gem5} simulation of CXL memory attached through the I/O bus, preserving an unmodified Linux kernel (v6.14+), drivers and software stack while capturing realistic timing, page-interleaving, and cache‑level interactions, accounting for the memory heterogeneity introduced by CXL memory expansion. 
Moreover, it facilitates benchmarking of existing and widespread programming models for CXL with realistic performance.

CXLRAMSim supports CXL memory simulation by modeling CXL Root Complex and the CXL.io/CXL.mem protocols to its fullest functionality, adhering to CXL2.0+ specifications (as supported by Linux v6.14). CXLRAMSim supports the CXL Command Line Interface (CXL-CLI) toolchain, thereby exposing the CXL memory in different ways to the OS. This makes CXLRAMSim a first in yet another domain, i.e., supporting prominent programming models for using CXL memory pool (memory interleaving, Flat memory mode, etc.). 

Our contributions are as follows:
\begin{itemize}
    \item We build a first full-system simulator for CXL based on gem5 that uses standard Linux kernel (v6.14) and driving, allowing the use of widespread existing programming models for using CXL memory. .
    \item Our setup enables the characterization interleaved accesses across CXL memory pool devices and system memory (DRAM), calibrated against a real hardware setup. CXLRAMSim v1.0 has been tested by exposing CXL memory as zNUMA node.
    \item We implement CXL protocols, namely, CXL.mem and CXL.io with packetization at the Root Complex and De-packetization at the CXL End point device. This way, the simulator implements architecturally correct  CXL extension for gem5. 
    \item 
    We implement a generic x86 BIOS model and CXL specific features needed to expose heterogeneity to the OS. This is a major development work for the x86 community that uses gem5, clearing roadblocks in modeling heterogeneous compute/memory systems simulated in gem5. 
\end{itemize}
\begin{figure}[t]
    \centering
    \includegraphics[
        width=\linewidth,
        trim=0 70mm 0 0,
        clip
    ]{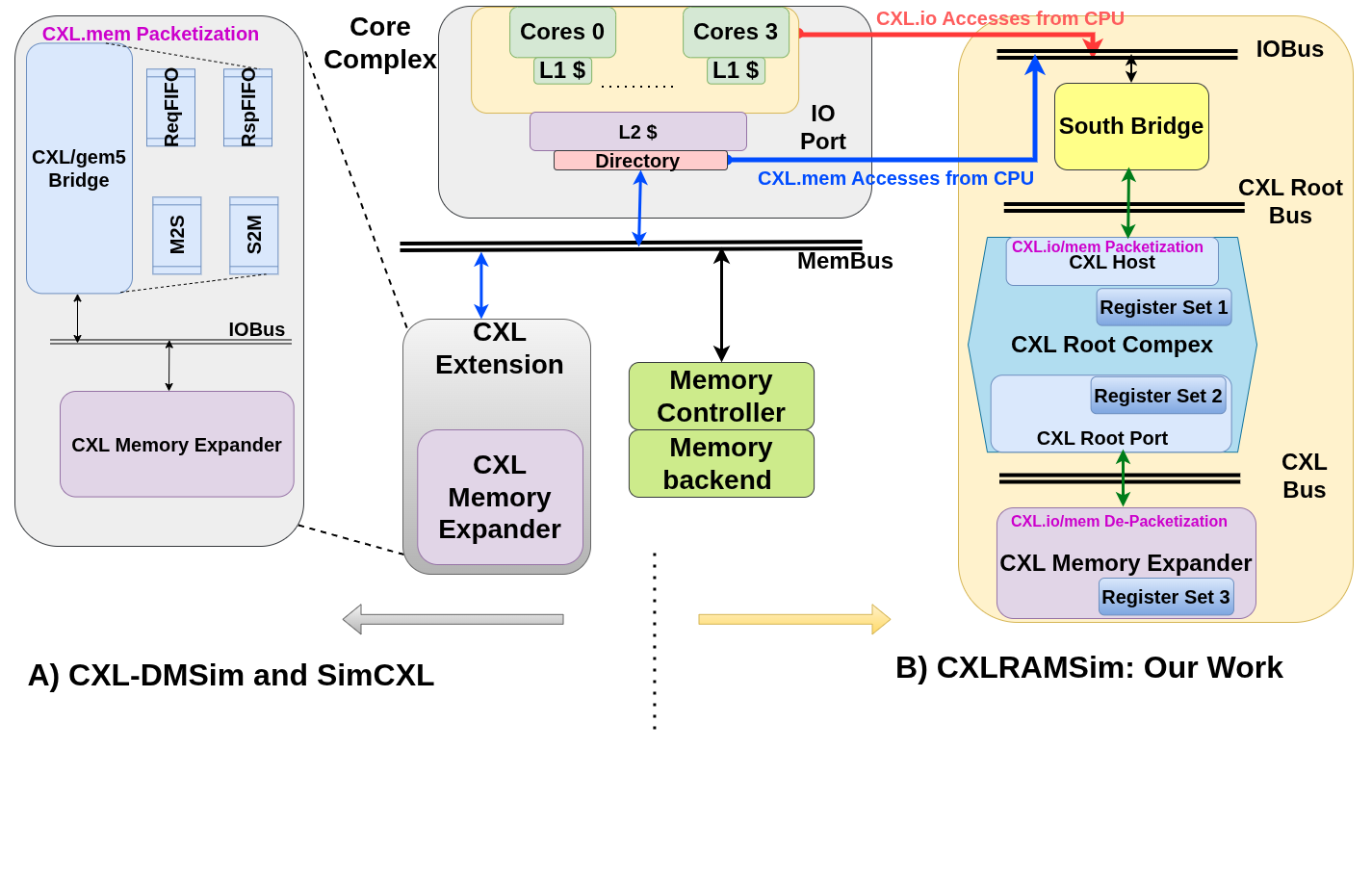}
    \caption{Full system simulators implementing CXL extension A) CXLDMSim and SimCXL on \textbf{MemBus} \cite{TCAD}, \cite{HPCA} and B) CXLRAMSim on \textbf{IOBus}.}
    \label{fig:CXLstack}
\end{figure}
We plan to open-source and integrate our work into gem5 mainstream repository. With our work, gem5 can now simulate and support NUMA systems, PCIe devices with full capabilities, multiple root complexes for PCIe/CXL/PCI (legacy) devices with entire PCIe hierarchy (Root Ports, Bridges and Buses).  

\section{Background}
\subsection{Related Work}
A plethora of simulation frameworks claim cycle accurate simulation of CXL devices. The most noteworthy among them are CXL-DMSim~\cite{TCAD} and SimCXL~\cite{HPCA}. However, these simulators do not even adhere to CXL 1.0 specifications for its CXL.io implementation, since they enumerate the CXL device as a PCI Memory Controller (not even as PCIe). Further, they have unrealistic interfaces as depicted in Figure \ref{fig:CXLstack}, connecting CXL devices to the Memory Bus (Membus) instead of to the IOBus. \textbf{This is akin to connecting a CXL memory on the DIMM slots of the motherboard, instead of on PCIe slots}. These architecturally incorrect interfaces are due to the lack of support for describing PCIe/CXL Hierarchy in the BIOS of the simulated system. Subsequently, the CXL memory is forced to be exposed as regular system DRAM,
resulting in the need to patch the Linux kernel and drivers\cite{TCAD}\cite{HPCA}. It has been seen that the simulators\cite{TECS} can achieve an accuracy of 2-10\% (for a chosen set of benchmarks) by tuning the parameters of the CPU model and load/store queues. The simulators insert queues (ReqFIFO, RespFIFO, etc.) in the modeled CXL extension to achieve even better accuracy. 
Moreover, these unrealistic interfaces make the simulator micro-architecture and statistics far from reality. Hence, the simulator can neither be used to identify nor can it be used to propose micro-architectural innovations to divest the system of the bottlenecks. Moreover, simulators \cite{TCAD}\cite{HPCA} propagate the same issues for other CXL device (Type-1 and Type-2 Accelerator) models and keep themselves devoid of the CXL specific enhancements in Linux kernel.
Other simulation frameworks are far more premature in simulating CXL memory devices in full system simulation \cite{TCAD}. 
\section{Simulation Framework}
\begin{figure*}[!t]
    \centering
    \includegraphics[width=0.6\linewidth]{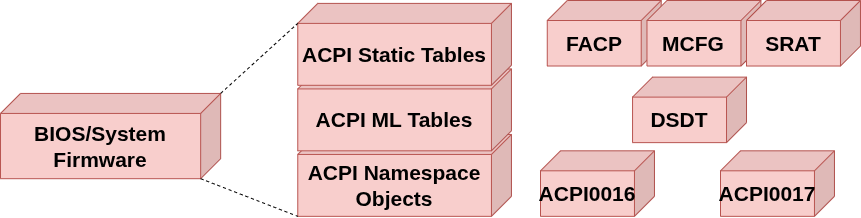}\\
    \caption{Modeled X86 Bios in gem5 to support CXL2.0 devices}
    \label{fig:CXL BIOS}
\end{figure*}
\begin{figure*}[b]
    \includegraphics[width=\linewidth]{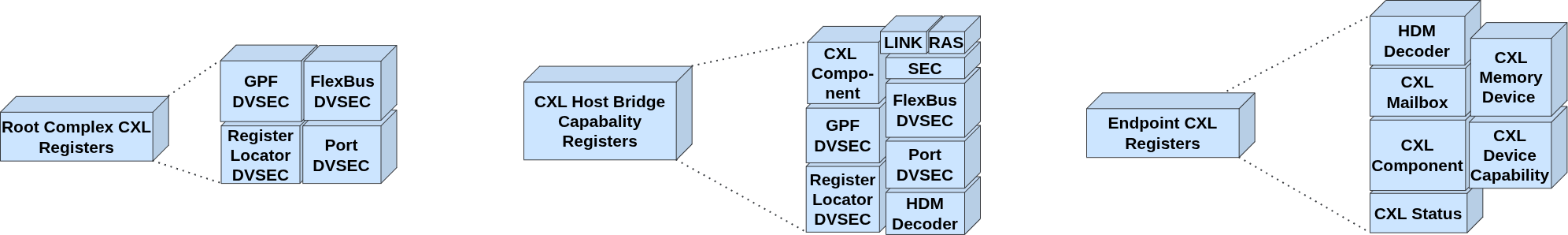}\\
    \caption{Modeled CXL Registers adhering to CXL2.0 specifications}
    \label{fig:Registers}
\end{figure*}
We build CXL Memory model on top of gem5 v25, with linux kernel 6.14v and ubuntu 24.04 LTS disk image. The simulator is also extended to support CXL-CLI to manage CXL memory device. The Kernel and drivers are used without modification by modeling BIOS of X86 systems in gem5, and is described below.
\subsection{Firmware Model}
The BIOS of gem5 for x86 comprises of three parts: \textit{E820} table entries describing the physical memory address map, \textit{ACPI} tables comprising of\textit{ RSDP} (Root System Description Pointer), \textit{MADT} (Multiple APIC Description Table), and Intel Multi-processor Configuration Table describing the local APIC base entry. This current BIOS is bare-minimum and suffice to boot OS (with Linux kernel). It can not support heterogeneous compute/memory systems including CXL devices. 

The CXLRAMSim fills this gap as it extends the gem5 to support modeling of heterogeneous architectures with x86 CPUs. We model \textit{MCFG} (Memory-mapped configuration) ACPI table to extend support for PCIE/CXL interfaces. The MCFG table allows the OS to discover the extra 4KiB configuration space of PCIe devices. The modeled BIOS is also extended to support ACPI Machine Language (ACPI ML) Interpreter, thereby enabling the guest OS to parse dynamic ACPI tables such as Differentiated System Description Table (DSDT), essential in describing the compute and memory heterogeneity to the OS. The DSDT table defines the topology of the modeled hardware and exposes it to the guest OS. It is critical in extending Support for CXL devices and creation of CXL hierarchy. It exposes the presence of CXL Host Bridge and Endpoint device via ACPI namespace objects to the OS. Moreover, the DSDT table defines the MMIO windows present in the hardware that can be used by the OS to memory map Base Address Registers (BARs) of PCIe/CXL devices.  
Finally, the parsing of DSDT tables invokes the OS to look for CXL Early Discovery Table (CEDT), another ACPI table, that registers the base address of the CXL Memory device when hot-plugged. We implement CXL Early Discovery Table (CEDT) that enumerates the CXL Memory device and exposes it as CPU-less NUMA node. The System Resource Affinity Table (SRAT) is modeled to define the affinity between CPU and Memory complexes. This is essential in modeling heterogeneous systems with CXL devices. 
In context of CXL, the SRAT and CEDT together define a CPU-less NUMA node for CXL memory, thereby, enabling support for zNUMA programming model (a prominent programming model) for CXL memory. The modeled BIOS for x86 adheres to CXL3.0 specifications and UEFI 2.10+ specification for CXL. A hihg level view of our contribution to the x86 Bios model in gem5 id depicted in Figure \ref{fig:CXL BIOS}

\subsection{CXL Hardware Model}
\subsubsection{CXL.io protocol}
\begin{figure*}[t]
    \centering
    \includegraphics[width=1\linewidth]{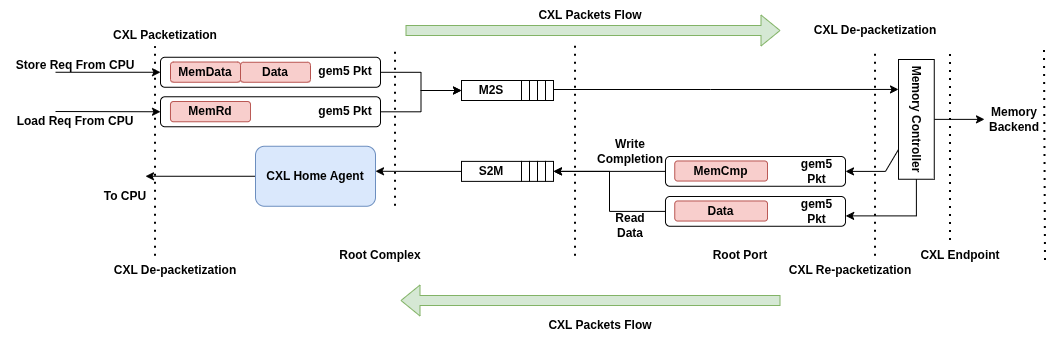}
    \caption{CXL.mem protocol Transaction Layer implementation in gem5}
    \label{fig:CXLmemprotocol}
\end{figure*}
At present, there is no full-system simulator with architecturally correct interfaces for PCIe/CXL cards. A fundamental entity to extend support for PCIe/CXL cards is a Root Complex which converts the host PCIe/CXL load/store requests to their corresponding protocol.  Moreover, a (CXL) Root Complex enables owning of resource bus required for describing PCIe/CXL hierarchy comprising of PCI(e)-PCI(e) bridge/ Root Port and PCIe/CXL End point device.  We model these entities with all required registers adhering to CXL 2.0+ specifications. 
Three sets of CXL specific (configuration space) registers are modeled as depicted in Figure \ref{fig:CXLstack}. Figure \ref{fig:Registers} further elaborates on the registers comprising of the three Register sets. \textit{ Set 1 }comprises of the Root Complex registers , namely,  Device Vendor Specific (DVSEC)  - GPF, Flexbus, Port and Register Locator Register. These registers are essential in exposing a CXL Root Complex with corresponding capabilities for the Linux drivers to bind to the CXL Root Complex.  \textit{Set 2} comprises of  the Host Bridge Registers, namely,  Link, RAS, Security (SEC), Component Registers and HDM Decoders registers that specify the address and size of the CXL devices connected beneath it. These registers are necessary to enumerate the Root Ports of the downstream CXL endpoint device. Lastly,\textit{ Set 3 }has the Mailbox and Status registers to enable user interaction of the modeled CXL device. Other Registers include the CXL Component and Device Capability Registers with implementing their specified functions as per CXL Specifications \cite{CXL30Spec}. 

With the implementation of x86 Bios and the above mentioned Register sets adhering to CXL2.0+ specifications, we fully implement the CXL.io protocol with its functionality. This also includes the support for CXL Command Line (CXL-CLI) toolchain (including \emph{NDCTL}~\cite{ndctl_user_guide}) needed for onlining the CXL memory post OS boot. This in conjunction of \emph{Numactl}~\cite{numactl} is used to ``online" and expose the CXL memory as CPU-less NUMA node to the OS. This has been made possible with the implementation of "Doorbell mechanism" (common to PCIe device), wherein the Status Registers is used by the host to probe the device status. Apart from the benefits of standard interrupt mechanism of interrupting CPU execution, this also allows user to interact from user space CXL-CLI. 
\subsubsection{CXL.mem protocol}
We model the Transaction layer of the CXL.mem protocol with packetization at the Root Complex and De-packetization at the CXL Endpoint Device as depicted in Figure \ref{fig:CXLmemprotocol}. We implement the M2S and S2M channels with CXL opcodes integrated into the packet headers. The CXL.mem protocol supports contention modeling via implementation of M2S and S2M. These separate the memory requests according to CXL.mem specifications, in the following manner:
\begin{itemize}
    \item M2S Request: We implement Opcode for Memory Read Requests from CPU (these are Load Requests).
    \item M2S Request with Data: We implement the Opcode for Memory Write  Requests from CPU (these are Store Requests).
    \item S2M No Data Response (NDR): We implement the Opcode for completion of Write Requests. This signifies that the CXL memory backend has updated the data with the Store request from the CPU.
    \item S2M Data Response (DRS): We implement the Opcode for making Memory Read data available to the CPU.  
\end{itemize}
\textbf{We expose the latency of the CXL packetization and De-packetization, CXL buses, etc. at the Python-level in gem5, thereby making it convenient for users to calibrate these latency with the actual hardware. }

\section{Experiments}
The CXL Memory with user defined size is ``onlined" and assigned to zNUMA (CPU-less) node  with user space libraries (NDCTL). Since we adhere to CXL2.0 specifications, the simulator can support creation of Mulitple Logic Devices (MLD). However, we restrict the focus of work to Single Logic Devices (SLDs) with MLDs to be taken up later.
\begin{figure*}[t]
    \centering
    \includegraphics[width=0.8\linewidth]{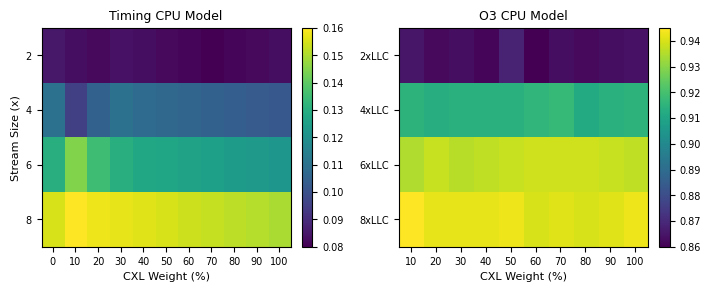}
    \caption{L2 Miss rate for Stream micro-benchmark}
    \label{fig:heatmaps}
\end{figure*}
\begin{table}[t]
\centering
\caption{Simulation Configuration}
\label{tab:sim_specs}
\begin{tabular}{ll}
\toprule
\textbf{Component} & \textbf{Specification} \\
\midrule
CPU Models & In-order, Out-of-Order \\
Cores & Up to 4 cores (x86 ISA) \\
Cache Coherence & MESI (Two-level, Directory-based) \\
System Memory & Configurable (Unbounded) \\
CXL Memory & Configurable Extension (Unbounded) \\
\bottomrule
\end{tabular}
\end{table}
The user can specify the size of memory assigned to zNUMA node, the rest of the memory in the modeled CXL card goes into the same NUMA node as System Memory. The later is called as Flat Memory mode as the OS perceives the CXL memory and System Memory as one big contiguous memory region. CXLRAMSim can support multiple programming models, including the Intel's Unified Memory Framework~\cite{oneapi_umf_intro}, Samsung's Scalable Memory Development Kit (SMDK)~\cite{SMDK}, SkHynix's Heterogeneous Memory Software Development Kit (HMSDK)~\cite{hmsdk_capacity_expansion}, etc., all of which expose the CXL device as a zNUMA node using CXL-CLI. 

Further, we characterize the CXL memory models with specifications described in the Table \ref{tab:sim_specs}. We execute Steam~\cite{intel_memory_bandwidth_benchmarks} micro-benchmarks, standard for System-level memory device characterization with MESI Two level cache coherence protocol. We execute stream with 2,4,6 and 8 times the size of L2 cache, thereby maximizing the stress on CXL memory. Also, we vary the OS managed page interleaving ratios between System memory and CXL memory, thereby proving that CXL memory models can handle few GiB of memory footprints. The Figure \ref{fig:heatmaps} depicts Last-Level Cache (LLC) missed with the Timing and Out-of-Order (O3) CPU models executing the Stream benchmarks. 

\section{Conclusion}
We provide a first ready-to-use full-system simulation framework for CXL memory cards with architecturally correct interfaces, supporting the most prominent programming models with utmost micro-architectural details on both host and device side. The bandwidth-latency characterstics of the CXL memory is highly vendor specific. Hence, we provide a user-friendly mechanism to calibrate the latency of the CXL interconnects to match the latency/bandwidth of the actual CXL memory.  We extend the cache coherence directory to incorporate the CXL memory into the coherence fabric of the CPU. We plan to release CXLRAMSim v2.0 with CXL Switches and CXLRAMSim v3.0 with Processing-Near-Memory over CXL memory. The release v3.0 would be a step towards supporting CXL Type-2 and CXL-Type 1 devices, planned to feature in v4.0.


\end{document}